\documentclass[reqno,draft]{amsart}
\usepackage{multicol, color}
\newcommand{\rem}[1]{}

\theoremstyle{plain}
\newtheorem{lemma}{Lemma}
\newtheorem{theorem}[lemma]{Theorem}

\newtheorem{proposition}[lemma]{Proposition}
\newtheorem{definition}[lemma]{Definition}
\theoremstyle{remark}
\newtheorem{remark}{Remark}

\newcommand*{\ang}[1]{\left\langle #1 \right\rangle}

\def\ee{\epsilon}
\def\aa{\alpha}

\def\ss{\sigma}

\def\la{\lambda}
\def\Dd{\Delta}

\def\Om{\Omega}

\def\pp{\partial}

\begin{document}
\title[Clark$-\alpha$ Model]
{On the Clark$-\alpha$ model of turbulence: \\
global regularity and
long--time dynamics}
\date{August 4, 2004}
\thanks{{\bf Submitted to:}\textit{Journal of Turbulence}}
\author[C. Cao]{Chongsheng Cao}
\address[C. Cao]
{Department of Mathematics  \\
University of Nebraska-Lincoln  \\
Lincoln, NE 68588-0323, USA\\
{\bf ALSO}  \\
Department of Mathematics \\
Florida International University \\
Miami, FL 33199, USA}
\email{ccao@math.unl.edu}
\author[D.D. Holm]{Darryl D. Holm}
\address[D.D. Holm]
{ T-7 \\
MS B284  \\
Los Alamos National Laboratory \\
Los Alamos, NM 87545, USA \\
{\bf ALSO}  \\
Department of Mathematics \\
Imperial College London, SW7 2AZ, UK
}
\email{dholm@lanl.gov and d.holm@imperial.ac.uk}
\author[E.S. Titi]{Edriss S. Titi}
\address[E.S. Titi]
{Department of Mathematics \\
and  Department of Mechanical and  Aerospace Engineering \\
University of California \\
Irvine, CA  92697-3875, USA \\
{\bf ALSO}  \\
Department of Computer Science and Applied Mathematics \\
Weizmann Institute of Science  \\
Rehovot 76100, Israel}
\email{etiti@math.uci.edu and edriss.titi@weizmann.ac.il}

\begin{abstract}
In this paper we study a well-known three--dimensional turbulence
model, the filtered Clark model, or Clark$-\alpha$ model {\bf
\cite{CFR79}}. This is Large Eddy Simulation (LES)
tensor-diffusivity model of turbulent flows with an additional spatial
filter of width $\alpha$. We show the global well-posedness of
this model with constant Navier-Stokes (eddy) viscosity. Moreover, we
establish the existence of a finite dimensional global attractor
for this dissipative evolution system, and we provide an anaytical
estimate for its fractal and Hausdorff dimensions. Our estimate is
proportional to $\left( L/l_d \right)^3$, where $L$ is the
integral spatial scale and $l_d$ is the viscous dissipation length
scale. This explicit bound is consistent with the physical estimate for
the number of degrees of freedom based on heuristic arguments. Using
semi-rigorous physical arguments we show that the inertial range of the
energy spectrum for the Clark-$\aa$ model has the usual $k^{-5/3}$
Kolmogorov power law for  wave numbers $k\aa \ll 1$ and $k^{-3}$ decay
power law for $k\aa \gg 1.$ This is evidence that the Clark$-\alpha$ model
parameterizes efficiently the large wave numbers within the inertial
range, $k\aa \gg 1$, so that they contain much less translational kinetic
energy than their counterparts in the Navier-Stokes equations.

\end{abstract}
\maketitle
\section{Introduction}   \label{S-1}
This paper is devoted to the mathematical analysis of a Large Eddy
Simulation (LES) model of turbulent flows with explicit filtering.
We consider the ``filtered Clark model,'' aka 
``Clark$-\alpha$ model''~\cite{CFR79}.
This is a nonlinear LES model of turbulence consisting of
the ``tensor-diffusivity model'' of Leonard \cite{Le1974}, filtered by
inversion of the Helmholtz operator with width  $\alpha$. As pointed out
in Winckelmans et al. \cite{WWVJ2001}, this model is generic: For all
regular symmetric filters that have a nonzero second moment, this form is
found as the first term of the reconstruction series for the
filtered-scale stress.  Thus, the Clark$-\alpha$ model we consider is
generic and corresponds to significant, yet perhaps incomplete,
reconstruction of the filtered-scale stress.
Such an explicitly filtered tensor-diffusivity model was used by Vreman et
al. \cite{VrGeKu1996,VrGeKu1997} with good success, in combination with a
dynamic Smagorinsky term as a ``nonlinear mixed model.'' In their work,
the Smagorinsky term was posed to model the truncation effects of
subgrid-scale stress and incomplete reconstruction of the filtered-scale
stress. Vreman et al. \cite{VrGeKu1996,VrGeKu1997} referred to their model
as ``a dynamic version of the mixed Clark model,'' while Winckelmans et al.
\cite{WWVJ2001} wrote that they preferred the term ``tensor-diffusivity
model'' for such models. The results of \cite{WWVJ2001} supported the view
that the explicitly filtered tensor-diffusivity (or, Clark) model
may well suffice for practical reconstruction of the filtered-scale stress,
{\it without} introducing a Smagorinsky term.

Section \ref{Def-1} is devoted to the precise definition of the
Clark$-\alpha$ model and to discussions of the properties of its
nonlinear terms.  Section \ref{ExitUnic-sec} establishes the main
results of global (in time) regularity for the Clark$-\alpha$
model. That is, global existence, uniqueness and continuous
dependence of the initial data of the solutions to this $3-D$
model of turbulence. Section \ref{S-A} shows that the
``Clark$-\alpha$ turbulence'' model possesses a global attractor
and gives an upper bound on its Hausdorff and fractal dimensions.
This upper bound is discovered to scale as $(L/l_d)^3$ -- the cube
of the integral scale $L$ divided by the dissipation length
$l_d=(\nu^3/\epsilon)^{1/4}$. Finally, the
translational kinetic energy spectrum $E(k)$ for the
Helmholtz-filtered Clark$-\alpha$ model is also shown in section
\ref{S-A} to pass from $k^{-5/3}$ for smaller wave numbers
($k\alpha \ll 1$), to $k^{-3}$ for larger wave numbers ($k\alpha
\gg 1$). It is worth mentioning that the results presented here
are similar to those already established for the
Navier--Stokes--$\aa$ (NS--$\aa$) model {\bf \cite{FHTM},
\cite{FHTP}} (also known as the three dimensional viscous
Camassa--Holm equations, or the Lagrangian averaged NS--$\aa$ model).
Furthermore, when viewed as a closure model of turbulence for the
Reynolds averaged equations in turbulent channels and pipes the
Clark--$\aa$ model gives exactly the same reduced equations in
channels and pipes as the NS--$\aa$ model. Comparing the solutions
of those reduced equations in pipes and channels to the empirical
data has already given excellent agreement (see, {\bf
\cite{CH98}--\cite{CH01}}.) Therefore, in this regard, one can
assert that the Clark--$\aa$ model is as successful as the
NS--$\aa$ sub--grid model of turbulence.

\section{Preliminaries and Notations}   \label{Def-1}
Let $\Om = [0, 2\pi L]^3$. The so-called Clark$-\alpha$ turbulence model
(cf. {\bf \cite{CFR79}})
of viscous incompressible flows in a domain $\Om$
subject to the periodic boundary condition reads:
\begin{eqnarray}
&&\hskip-.8in
\partial_t u  - \nu   \Dd u + (u \cdot \nabla ) u + \nabla p -
\nabla \cdot ( \mathcal{H}^{-1} ( \aa^2 \nabla u \cdot \nabla u^T))=f,
 \label{CEQ-1}  \\
&&\hskip-.8in
\nabla \cdot u= 0,  \label{CEQ-2}  \\
&&\hskip-.8in
u(0) =u_0,    \label{CEQ-3}
\end{eqnarray}
where, $u$ represents the unknown ``averaged/filtered" fluid
velocity vector, and $p$ is the unknown ``averaged/filtered"
pressure scalar; $\nu
>0$ is the constant kinematic (eddy) viscosity, $\aa$ is a length scale
parameter which represents the width of the filter. The body
forcing term, $f$, which is time independent, and the initial
velocity, $u_0$, are given. The operator $\mathcal{H}$ is the
Helmholtz operator, defined as
\[
\mathcal{H}u = u-\aa^2 \Dd u, \quad \mbox{subject to periodic
boundary condition},
\]
and the tensor $(\nabla u \cdot \nabla u^T)$ is given by
\[
(\nabla u \cdot \nabla u^T)_{ij} = \nabla u_i \cdot \nabla u_j
=\sum_{k=1}^{3} (\pp_{x_k} u_i)\; ( \pp_{x_k} u_j).
\]
{F}rom now on, the Einstein's summation convention will be used,
e.g.,
\[
(\pp_{x_k} u_i ) (\pp_{x_k} u_j) =
\sum_{k=1}^{3} (\pp_{x_k} u_i) (\pp_{x_k} u_j).
\]
Let $v=\mathcal{H}u = u-\aa^2 \Dd u$. Then, the above system is
equivalent to
\begin{eqnarray}
&&\hskip-.8in
\partial_t v  - \nu   \Dd v + (u \cdot \nabla ) v + (v \cdot \nabla ) u
- (u \cdot \nabla ) u -\aa^2
\nabla \cdot (  \nabla u \cdot \nabla u^T )+ \nabla q =g,
 \label{CCEQ-1}  \\
&&\hskip-.8in
\nabla \cdot u=\nabla \cdot v= 0,  \label{CCEQ-2}  \\
&&\hskip-.8in
v(0) =v_0 = \mathcal{H} u_0,    \label{CCEQ-3}
\end{eqnarray}
where, $q=\mathcal{H} p$ and $g=\mathcal{H} f$.

\subsection{Notations}
Let $L^p(\Om)$ and $H^m(\Om)$ denote the usual $L^p$ Lebesgue
spaces and Sobolev spaces, respectively (cf. {\bf \cite{AR75}}).
We will denote by $| \cdot |$ the $L^2-$ norm, and $(\cdot,
\cdot)$ the $L^2-$ inner product. Let $\mathcal{F}$ be the
function space which consists of all the vector trigonometric
polynomials. We set
\[
\mathcal{V}= \{ \phi \in \mathcal{F}:
\nabla \cdot \phi =0 \quad \& \quad \int \phi =0 \}.
\]
The spaces $H, V_1$ and $V_2$ will denote the closures of
${\mathcal{V}}$ in $L^2, H^1$ and $H^2$ respectively. Let
$P_{\ss}:L^2 \rightarrow H$, be the orthogonal projection, and let
$A=-P_{\ss} \Dd$ be the Stokes operator subject to the periodic
boundary conditions. It is well known that $A^{-1}$ is a
self-adjoint positive compact operator in $H$ and ${\mathcal{D}}
(A) = V_2$ (cf. {\bf \cite{CF88}}, {\bf \cite{GA94}}, {\bf
\cite{LADY}}, {\bf \cite{TT84}}). Let $0 < \la_1=L^{-2} \leq \la_2
\leq \cdots $ be the eigenvalues of $A$, repeated according to
their multiplicities. It is well-known that a constant $C_1 >0$
exists such that
\begin{eqnarray}
&&\hskip-.8in \frac{j^{2/3}}{C_1} \leq \frac{\la_j}{\la_1} \leq
C_1 j^{2/3},  \label{LL}
\end{eqnarray}
for $j=1, 2, \cdots, $ (see, e.g., {\bf \cite{CF88}}).

\vskip0.1in For $w_1, w_2 \in \mathcal{V}$, we define
\[
B(w_1, w_2)  =P_{\ss} \left((w_1\cdot \nabla )w_2\right).
\]
The bilinear form $B$ has the following properties (cf.
\cite{CF88}, \cite{TT84}).
\begin{proposition}
The bilinear form $B$ can be extended as a continuous map
$B:V_1\times V_1\rightarrow V_1^{\prime},$ where $V_1^{\prime}$
is the dual space of $V_1$. Furthermore,
\[
\ang{B(w_1,w_2),w_3}_{V_1^{\prime}} = -\ang{B(w_1,w_3),w_2}_{V_1^{\prime}}
\]
for every $w_1, w_2, w_3 \in V_1.$
\end{proposition}
\begin{proposition}
Let $u\in H^3\cap V_1, v=\mathcal{H} u =u-\aa^2 \Dd u$. Then, the bilinear
operators
$B(u,v), B(v,u)$  and 
$\displaystyle{\sum_{j=1}^{3}} B(\pp_{x_j} u, \pp_{x_j} u)$ are
well defined with values in $V_1^{\prime}$.
\end{proposition}

The following properties can be established by integration by parts.
\begin{proposition} \label{P-3}
Let $u\in H^3\cap V_1, v=\mathcal{H} u =u-\aa^2 \Dd u$. Then,
\begin{enumerate}
\item $\ang{B(u,u),u}_{V_1^{\prime}} = 0,$
\item $\ang{B(v,u),u}_{V_1^{\prime}} = 0,$
\item $\ang{B(u,v),u}_{V_1^{\prime}}
-\aa^2 \displaystyle{\sum_{j=1}^{3}}
\ang{B(\pp_{x_j} u, \pp_{x_j} u), u}_{V_1^{\prime}}  = 0.$
\end{enumerate}
\end{proposition}
As for the case of the Navier--Stokes equations 
(cf.~\cite{CF88},~\cite{TT84})
by applying the operator $P_{\ss}$, we get the following system which is
equivalent to the system (\ref{CEQ-1})--(\ref{CEQ-3})
\begin{eqnarray}
&&\hskip-.8in
\frac{d u}{d t}  + \nu   A u + B(u, u)
+ \aa^2 \mathcal{H}^{-1} B(\pp_{x_j} u, \pp_{x_j} u)
= f,
 \label{EQ-1}  \\
&&\hskip-.8in
u(0) =u_0,    \label{EQ-3}
\end{eqnarray}
and
the following system which is equivalent
to the system (\ref{CCEQ-1})--(\ref{CCEQ-3})
\begin{eqnarray}
&&\hskip-.8in
\frac{d v}{d t}  + \nu   A v + B(u, v)
+B(v, u)- B(u, u) - \aa^2 B(\pp_{x_j} u, \pp_{x_j} u)
= g,
 \label{NEQ-1}  \\
&&\hskip-.8in
v(0) =v_0.    \label{NEQ-3}
\end{eqnarray}

\begin{definition} \label{D-1}
\thinspace Let $T$ be any fixed positive number and $f \in V_1$.
\thinspace We call $u$ a regular solution of {\em
(\ref{EQ-1})--(\ref{EQ-3})} as an equation in $V_1$ on $[0,T]$ if
$u$ satisfies the system {\em (\ref{EQ-1})--(\ref{EQ-3})} and
\begin{eqnarray*}
&&\hskip-1.0in
u   \in C_w([0,T], V_2) \cap L^2([0,T], H^3 (\Om)), \quad  \mbox{and} \\
&&\hskip-1.0in
\frac{d u}{d t}     \in L^2 ([0,T], V_1).
\end{eqnarray*}
Here $C_w([0,T], V_2)$ is the functional space of all weakly
continuous functions from $[0,T]$ to $V_2$.
\end{definition}
For convenience, we recall the following Sobolev and
Ladyzhenskaya's inequalities in $\mathbb{R}^3$
\begin{eqnarray}
&&\hskip-.68in
\| u \|_{L^3} \leq C_0 \| u \|_{L^2}^{1/2} \| u \|_{H^1}^{1/2},  \label{SI-1}\\
&&\hskip-.68in
\| u \|_{L^4} \leq C_0 \| u \|_{L^2}^{1/4} \| u \|_{H^1}^{3/4},  \label{SI-2}\\
&&\hskip-.68in
\| u \|_{L^6} \leq C_0 \| u \|_{H^1}, \label{SI-3}
\end{eqnarray}
for every $u\in V_1,$ and the Agmon's  inequlity in $\mathbb{R}^3$
\begin{eqnarray}
&&\hskip-.68in
\| u \|_{L^{\infty}} \leq C_0 \| u \|_{H^1}^{1/2} \| u \|_{H^2}^{1/2},  \label{AI}
\end{eqnarray}
for every $u\in V_2$. Here $C_0$ is a universal constant. Also,
notice that
\begin{eqnarray}
&&\hskip-.68in
| \mathcal{H} u |^2 \leq \left( \la_1^{-1} +\aa^2 \right)
( |\nabla u|^2+\aa^2 |\Dd u|^2).   \label{LLL}
\end{eqnarray}

For this model the quantity
\[
(u,v) = |u|^2 + \aa^2 |\nabla u|^2
\]
will be called the energy of the system, while the
quantity  $(u,u)=|u|^2$ will be called the translational energy of the system. 

\section{Existence and Uniqueness} \label{ExitUnic-sec}
\subsection{$H^1$ estimate} The estimates presented here are formal.
They can be made rigorous by establishing them first for a
Galerkin approximating scheme and then passing to the limit by
using Aubin compactness theorem (see, e.g., {\bf \cite{CF88},
\cite{TT84}} and {\bf \cite{TT88}}). By taking the $V_1^{\prime}$
action of equation (\ref{EQ-1}) with $v=\mathcal{H} u$ and using
Proposition \ref{P-3} and Lemma 1.2 in Chapter III of {\bf
\cite{TT84}}, one finds the energy balance satisfies:
\begin{eqnarray}
&&\hskip-.68in \frac{1}{2} \frac{d (|u|^2+ \aa^2|\nabla u|^2)}{dt}
+ \nu ( |\nabla u|^2+\aa^2 |\Dd u|^2) + \ang{ B(u, u),
\mathcal{H}u }_{V_1^{\prime}} - \aa^2 \ang{ B(\pp_{x_j} u,
\pp_{x_j} u), u}_{V_1^{\prime}} = \ang{ f,
\mathcal{H}u}_{V_1^{\prime}}. \label{L-1}
\end{eqnarray}
By using part (iii) of Proposition \ref{P-3} we obtain
\begin{eqnarray}
&&\hskip-.68in
\frac{1}{2} \frac{d (|u|^2+ \aa^2|\nabla u|^2)}{dt}
+ \nu ( |\nabla u|^2+\aa^2 |\Dd u|^2)
= (f, \mathcal{H}u)  \\
&&\hskip-.68in
\leq |f|\; |u| + \aa^2 |\nabla f| |\nabla u| \\
&&\hskip-.68in
\leq \frac{1}{2\nu \la_1} (|f|^2 + \aa^2 |\nabla f|^2)
+ \frac{\nu \la_1}{2} ( |u|^2+\aa^2 |\nabla u|^2)\\
&&\hskip-.68in
\leq \frac{1}{2\nu\la_1} (|f|^2 + \aa^2 |\nabla f|^2)
+ \frac{\nu}{2} ( |\nabla u|^2+\aa^2 |\Dd u|^2).
\end{eqnarray}
As a result, we reach
\begin{eqnarray}
&&\hskip-.68in
\frac{d (|u|^2+ \aa^2|\nabla u|^2)}{dt}
+ \nu ( |\nabla u|^2+\aa^2 |\Dd u|^2)
\leq \frac{|f|^2 +\aa^2 |\nabla f|^2}{\nu\la_1}.
\label{V1}
\end{eqnarray}
Thanks to the Gr\"onwall inequality, we get
\begin{eqnarray}
&&\hskip-.68in
|u|^2+ \aa^2|\nabla u|^2 \leq K_1(\aa, \nu, t),
\label{K-1}
\end{eqnarray}
where
\begin{eqnarray}
K_1(\aa, \nu, t) =
 e^{- \nu \la_1 t} (|u_0|^2+ \aa^2|\nabla u_0|^2)
+\frac{ |f|^2 +\aa^2 |\nabla f|^2}{\nu^2 \la_1^2}.
\label{K1}
\end{eqnarray}
Moreover, by (\ref{V1}), for each $r>0$, we have
\begin{eqnarray}
&&\hskip-.68in
\nu \int_t^{t+r} ( |\nabla u|^2+\aa^2 |\Dd u|^2)
\leq \frac{(|f|^2 +\aa^2 |\nabla f|^2) r }{\nu\la_1}
+K_1(\aa, \nu, t).        \label{RT}
\end{eqnarray}
Consequently,
\begin{eqnarray}
&&\hskip-.68in
\int_0^{t} ( |\nabla u|^2+\aa^2 |\Dd u|^2)   \leq
\frac{(|f|^2 +\aa^2 |\nabla f|^2) t }{\nu^2\la_1} +
\frac{1}{\nu} \left(|u_0|^2+ \aa^2|\nabla u_0|^2\right).   \label{INT}
\end{eqnarray}

\subsection{$H^2$ estimate}
Again, the estimates presented here are formal. As above,
they can be made rigorous by establishing them first for a
Galerkin approximating scheme and then passing to the limit. By
taking the $V_1^{\prime}$ action of equation (\ref{NEQ-1}) with
$v=\mathcal{H} u$, one finds
\begin{eqnarray}
&&\hskip-.68in \frac{1}{2} \frac{d |v|^2}{dt} + \nu |\nabla v|^2 +
\ang{B(u,v)+ B(v,u)-B(u,u) - \aa^2 B(\pp_{x_j} u, \pp_{x_j} u),
v}_{V_1^{\prime}} = \ang{g, v}_{V_1^{\prime}}.
\end{eqnarray}
Notice that
\begin{eqnarray*}
&&\hskip-.68in \left| \ang{B(u,v)+B(v,u)-B(u,u)
- \aa^2 B(\pp_{x_j} u, \pp_{x_j} u), v}_{V_1^{\prime}}  \right| \\
&&\hskip-.68in =\left| \ang{B(v,u)-B(u,u)
- \aa^2 B(\pp_{x_j} u, \pp_{x_j} u), v}_{V_1^{\prime}} \right| \\
&&\hskip-.68in \leq |\nabla u| \; \|v\|_{L^4}^2 + |v| \;
\|u\|_{L^3} \|u\|_{L^6}
+ \aa^2 |\nabla v| \; \|\nabla u\|_{L^4}^2 \\
&&\hskip-.68in
\leq C_0^2 |\nabla u| \; |v|^{1/2} |\nabla v|^{3/2}
+ C_0^2 |v| \; |u|^{1/2} |\nabla u|^{3/2}
+C_0^2  \aa^2 |\nabla v| \; |\nabla u|^{1/2}  |\Dd u|^{3/2}.
\end{eqnarray*}
Also,
\begin{eqnarray*}
&&\hskip-.68in
\left| \ang{g,v}_{V_1^{\prime}}=\ang{\mathcal{H} f,v}_{V_1^{\prime}} \right|
= \left| \ang{f, v}_{V_1^{\prime}} -\aa^2 \ang{\Dd f, v}_{V_1^{\prime}}  \right| \\
&&\hskip-.68in \leq |f| \; |v| + \aa^2 |\nabla f| \; |\nabla v|.
\end{eqnarray*}
By Cauchy--Schwarz  and Young's inequalities we reach
\begin{eqnarray*}
&&\hskip-.68in
\frac{d |v|^2 }{dt}  + \frac{\nu}{4} |\nabla v|^2 \\
&&\hskip-.68in
\leq |f|^2 + |v|^2 + \aa^4 |\nabla f|^2
+ \frac{2C_0^8 |\nabla u |^{4} | v |^{2}}{\nu^3}
+8 C_0^2 |v| \; |u|^{1/2} |\nabla u|^{3/2}
+ \frac{8 C_0^4 \aa^4 | \nabla u | \;| \Dd u |^{3} }{ \nu }  \\
&&\hskip-.68in
\leq |f|^2 +\aa^4 |\nabla f|^2
+ \left[ 1+  \frac{2C_0^8 |\nabla u |^{4}}{\nu^3}
+ \frac{8C_0^2 \la_1^{1/2}  |u|^{1/2} | \nabla u |^{1/2} }{ \aa^2}
\frac{8C_0^4 | \nabla u | \;| \Dd u | }{ \nu }
\right] \; |v|^2.
\end{eqnarray*}
By (\ref{K-1}), (\ref{INT}),
and Gr\"onwall inequality we get
\begin{eqnarray}
&&\hskip-.68in |v(t)|^2 \leq K_2(\aa, \nu, t) \left( |v|^2(s)+ t(
|f|^2 +\aa^4 |\nabla f|^2 ) \right), \label{K-22}
\end{eqnarray}
where
\begin{eqnarray}
&&\hskip-.08in K_2(\aa, \nu, t) =\exp\left(  \left[ 1+
\frac{2C_0^8 (K_1(\aa, \nu, 0))^2 }{ \nu^3 \aa^4} + \frac{8C_0^2
\la_1^{3/4} (K_1(\aa, \nu, 0))^{1/2}}{ \aa^3} +\frac{8C_0^4 |f|^2
}{\nu^3 \aa^2 \la_1^{1/2} } \right] \; t \right.
\nonumber\\&&\hspace{2in}
\left. +\,\frac{8C_0^4 |u_0|^2+
\aa^2|\nabla u_0|^2}{\nu^2} \right). \label{K2}
\end{eqnarray}
In particular,
\begin{eqnarray}
&&\hskip-.68in
|v(t)|^2 \leq K_2(\aa, \nu, t) \left( |v_0|^2 +
t(|f|^2 +\aa^4 |\nabla f|^2) \right). \label{K-2}
\end{eqnarray}
Moreover,
\begin{eqnarray}
&&\hskip-.68in
\frac{\nu}{4} \int_0^{t} |\nabla v|^2 \leq
(|f|^2 +\aa^4 |\nabla f|^2) \; t
+K_2(\aa, \nu, t)\left( |v_0|^2 + t(|f|^2 +\aa^4 |\nabla f|^2) \right).   \label{INT-2}
\end{eqnarray}
These analytical estimates lead to the following.

\begin{theorem} \label{T}
Let $u_0 \in V_1$ and $f\in V_1$. Then there is a unique regular solution
to system (\ref{EQ-1})--(\ref{EQ-3}) for all $t>0$. Furthermore, this
solution depends continuously on the initial data in the sense that it
will specified in the proof below.
\end{theorem}
\begin{proof}
One can establish the existence of a regular solution by applying
the standard Galerkin approximation procedure together with the
{\it a priori} estimates (\ref{K-1}), (\ref{INT}), (\ref{K-2}) and
(\ref{INT-2}) (see, for example, {\bf \cite{CF88}}, {\bf
\cite{GA94}}, {\bf \cite{LADY}}, {\bf \cite{TT84}}). Here, we
shall only show the uniqueness of regular solution to the system
(\ref{EQ-1})--(\ref{EQ-3}) and the continuous dependence of solutions
on the initial data. Suppose that $u_1$ and $u_2$ are two
regular solutions to the system (\ref{EQ-1})--(\ref{EQ-3}), and
let $w=u_1-u_2$. Then the difference $w$ satisfies
\begin{eqnarray}
&&\hskip-.8in
\frac{dw}{dt}  + \nu A w +
 + B(w, u_1) + B(u_2, w) + \aa^2 \mathcal{H}^{-1} B(\pp_{x_j} w, \pp_{x_j} u_1)
+ \aa^2 \mathcal{H}^{-1} B(\pp_{x_j} u_2, \pp_{x_j} w) = 0,
 \label{UEQ-1}  \\
&&\hskip-.8in
w(0) =0.    \label{UEQ-3}
\end{eqnarray}
By taking the $V_1^{\prime}$ action of equation (\ref{UEQ-1}) with
$\mathcal{H} w$ and applying Lemma 1.2 in Chapter III of {\bf \cite{TT84}},
one finds
\begin{eqnarray*}
&&\hskip-.68in
\frac{1}{2} \frac{d (|w|^2+ \aa^2|\nabla w|^2)}{dt}
+ \nu ( |\nabla w|^2+\aa^2 |\Dd w|^2)  \\
&&\hskip-.68in =- \ang{ B(w, u_1) + B(u_2, w) + \aa^2
\mathcal{H}^{-1} B(\pp_{x_j} w, \pp_{x_j} u_1)
+ \aa^2 \mathcal{H}^{-1} B(\pp_{x_j} u_2, \pp_{x_j} w), \mathcal{H} w }_{V_1^{\prime}}  \\
&&\hskip-.68in \leq |\mathcal{H} w| (\|w\|_{L^{\infty}} |\nabla
u_1| + \|u_2\|_{L^{\infty}} |\nabla w|)
+ \aa^2 (|\nabla u_1|+|\nabla u_2|) \|\nabla w\|_{L^4}^2   \\
&&\hskip-.68in
\leq C_0 |\mathcal{H} w| (|\nabla w|^{1/2} |\Dd w|^{1/2}  |\nabla u_1|
+ |\Dd u_2| |\nabla w|)
+ \aa^2 C_0 (|\nabla u_1|+|\nabla u_2|) |\nabla w|^{1/2} |\Dd w|^{3/2}.
\end{eqnarray*}
By Cauchy--Schwarz inequality, we obtain
\begin{eqnarray*}
&&\hskip-.68in
\frac{1}{2} \frac{d (|w|^2+ \aa^2|\nabla w|^2)}{dt}
\leq C \left( |\Dd u_2|^2+ |\nabla u_1|^4+|\nabla u_2|^4\right)
|\nabla w|^2,
\end{eqnarray*}
where $C$ is a constant.
Thanks to the Gr\"onwall inequality, we get
\begin{eqnarray*}
&&\hskip-.68in |w(t)|^2+ \aa^2|\nabla w (t)|^2 \leq \exp \left(
\frac{C}{\aa^2} \int_0^t  \left( |\Dd u_2|^2 + |\nabla
u_1|^4+|\nabla u_2|^4 \right) \right) \left(|w(0)|^2+ \aa^2|\nabla
w(0)|^2\right).
\end{eqnarray*}
Thanks to (\ref{K-1}), (\ref{INT}),  (\ref{K-2})
and (\ref{INT-2}), the above implies continuous dependece on
initial data. In particular,
\[
w(t)=0   \quad \mbox{when} \;\; w(0) =0.
\]
Therefore, the regular solution is unique.
\end{proof}

\begin{remark} In order to be able to estimate the
dimension of the global attractor it is required that the solution
is differentiable, in the appropriate norms, with respect to the
initial data. Following similar energy methods to the ones
introduced above one can establish the required differentiability
(see, e.g., \cite{CF85}, \cite{CF88} and \cite{TT88}).
\end{remark}

\section{Global Attractors and Energy Spectra}   \label{S-A}
In this section we show the existence of the global attractor.
Moreover, we provide an upper bound to its fractal and Hausdorff
dimension.
Also, we consider the energy spectrum for the Clark$-\alpha$ model.
\subsection{Global Attractor}
Denote by $u(t) =S(t)u_0$
the solution of the system (\ref{EQ-1})--(\ref{EQ-3}) with initial
data $u_0$. As a result of Theorem \ref{T}, one can show that
\[
u(t) =S(t)u_0 \in V_1 \qquad \mbox{ for all} \quad
u_0 \in  V_1,  t \geq 0,
\]
and
\[
u(t) =S(t)u_0 \in V_2 \qquad
\mbox{ for all} \quad  u_0 \in V_2,
t \geq 0.
\]
Since, in this section,  we only consider the long time behavior of
solutions of the system (\ref{EQ-1})--(\ref{EQ-3}),  by (\ref{INT}), (\ref{K-22}) and
Theorem \ref{T}, we conclude that
$u(t) \in L^{\infty}_{\mbox{loc}}((0,S], V_2)$
for every $u(0) \in V_1$ and any $S>0$.
\begin{theorem} \label{T-A}
Suppose that $f \in V_1$. Then, there is a compact global
attractor $\mathcal{A} \subset  V_1$ for the system {\em
(\ref{EQ-1})--(\ref{EQ-3})}. Moreover, $\mathcal{A}$ has finite
Hausdorff and fractal dimensions.
\end{theorem}
\begin{proof}
First, let us show that there is an absorbing ball in
$V_1$ and $V_2.$
Let $u$ be the solution of the system
(\ref{EQ-1})--(\ref{EQ-3}) with initial data
$u_0 \in  V_1$ and
$|u_0|^2+ \aa^2|\nabla u_0|^2 \leq \rho.$
By  (\ref{V1}), we have
\begin{eqnarray*}
|u(t)|^2+ \aa^2|\nabla u(t)|^2
&\leq&  e^{- \nu \la_1 t} (|u_0|^2+ \aa^2|\nabla u_0|^2)
+\frac{ |f|^2 +\aa^2 |\nabla f|^2}{\nu^2 \la_1^2}   \\
&\leq& e^{-\nu \la_1  t} \rho^2
+\frac{ |f|^2 +\aa^2 |\nabla f|^2}{\nu^2 \la_1^2}.
\end{eqnarray*}
As a result of the above, when $t$ is large enough such that
\begin{eqnarray*}
&&\hskip-.68in
e^{-\nu \la_1  t} \rho^2 \leq
\frac{ |f|^2 +\aa^2 |\nabla f|^2}{\nu^2 \la_1^2},
\end{eqnarray*}
we have
\begin{eqnarray}
&&\hskip-.3in
|u(t)|^2+ \aa^2|\nabla u(t)|^2
\leq  R_a(\aa, \nu, f),   \label{R1}
\end{eqnarray}
where
\begin{eqnarray}
&&\hskip-.3in
R_a(\aa, \nu, f)=
2 \frac{ |f|^2 +\aa^2 |\nabla f|^2}{\nu^2 \la_1^2}.
\label{RA}
\end{eqnarray}
In particular,
\[
\limsup_{t \rightarrow \infty} (|u(t)|^2+ \aa^2|\nabla u(t)|^2)
\leq 2 \frac{ |f|^2 +\aa^2 |\nabla f|^2}{\nu^2 \la_1^2}.
\]
Therefore,  system (\ref{EQ-1})--(\ref{EQ-3}) has an absorbing ball
$\mathcal{B}$ in $V_1$ with radius $R_a(\aa, \nu, f)$.
\vskip0.1in
Next, we show that there is an absorbing ball in $V_2.$
First, by (\ref{R1}) and (\ref{RT})
we have
\begin{equation}
\int_t^{t+r} ( |\nabla u|^2+\aa^2 |\Dd u|^2)
\leq
\nu \int_t^{t+r} ( |\nabla u|^2+\aa^2 |\Dd u|^2)
\leq
 \frac{ |f|^2 +\aa^2 |\nabla f|^2 \left( 2+ r \nu \la_1
\right) }{\nu^3 \la_1^2}.         \label{RR}
\end{equation}
By  applying the uniform Gr\"onwall inequality (cf. for example,
{\bf \cite{TT88}}, p. 89) and (\ref{RR}) and (\ref{INT}), we
obtain,
  when $t$ is large enough,
\begin{equation}
| \mathcal{H} u(t)|^2 = | v(t)|^2 \leq R_v (r, \aa, \nu, f),   \label{R2}
\end{equation}
where
\begin{eqnarray}
&&\hskip-1.08in R_v (r,\aa, \nu,f) = \displaystyle{\exp \left(
\frac{C_0^{1/2}}{ \aa^2} \left( R_a(\aa, \nu, f) \right)^{1/2} +
\frac{ 2 r^{1/3}}{ \aa^{8/3} \nu^{4/3}} \left( \frac{ |f|^2 +\aa^2
|\nabla f|^2 \left( 2+ r \nu \la_1 \right) }{\nu^3 \la_1^2}
\right)^{2/3}
\right) }  \times  \nonumber  \\
&&  \times
\left( \frac{ \left(|f|^2 +\aa^2 |\nabla f|^2 \right) \left( 2+ r \nu \la_1
\right) }{\nu^3 \la_1^2 r}  +
\frac{r}{\nu} |\nabla f|^2 \right) ,  \label{RV}
\end{eqnarray}
and $r>0$ is fixed. Therefore, we conclude there is an absorbing
ball $\mathcal{B}$ in $V_2$ with radius $R_v (r, \aa, \nu, f)$.
Thanks to Rellich Lemma (see, e.g., {\bf \cite{AR75}}), the
operator $S(t)$ is a compact operator from $V_1$ to itself.
Following the standard procedure (cf., for example, {\bf
\cite{CF85}}, {\bf \cite{CF88}}, {\bf \cite{EFNT}, \cite{LADY91},
\cite{TT88}} for details), one can prove that there is a global
attractor
\[
\mathcal{A} =\cap_{s>0} (\cup_{t>s} S(t) \mathcal{B})  \subset V_2.
\]
Moreover, $\mathcal{A}$ is compact in $V_1$.

\vskip0.1in
By using the estimates (\ref{R1}) and (\ref{RR}), applying Lemma 4 and Lemma 5 in \cite{FHTM}, and
following the proof of Theorem 6 in \cite{FHTM} line by line,
we obtain the Hausdorff and fractal dimensions of the
attractor $\mathcal{A}$
\[
d_H (\mathcal{A}) \leq d_F (\mathcal{A}) \leq
C \max \left\{  \frac{ \left(|f|^2+\aa^2 |\nabla f|^2\right)^{2/3}}{\aa^{4/3} \nu^{8/3}
\la_1^{5/3}}, \frac{ \left(|f|^2+\aa^2 |\nabla f|^2\right)^{3/4}}{\aa^{3/4} \nu^{3}
\la_1^{3/2}}
\right\}.
\]
Moreover, following \cite{FHTM}, we define the mean rate of
dissipation of energy,
\begin{equation}
\ee=\displaystyle{\sup_{u_0 \in \mathcal{A}}}
\displaystyle{\limsup_{T\rightarrow\infty}} \frac{\nu}{T}
\int_0^{T} (\| \nabla u (t)\|_{L^2}^2 + \aa^2 \| \Dd u(t)
\|_{L^2}^2 ), \label{EEE}
\end{equation}
where $u(t)$ is the solution corresponding to the initial datum
$u_0$. By analogy, we define the Kolmogorov dissipation length
scale corresponding to the system (\ref{CEQ-1})--(\ref{CEQ-3}),
\[
l_d = \left( \frac{\nu^3}{\ee} \right)^{1/4}
\,.
\]
Hence, again, following the proof of Theorem 7 in \cite{FHTM}, we
have
\[
d_H (\mathcal{A}) \leq d_F (\mathcal{A}) \leq C \left( \frac{L}{\aa} \right)^{3/4}
\left( \frac{L}{l_d} \right)^{3}.
\]
We refer the reader to \cite{FHTM} for more details.
\end{proof}

\section{Spectral Scaling}
Following  \cite{FHTP} (see also {\bf \cite{FOIAS}} and {\bf
\cite{FMRT}}), we set
\begin{eqnarray*}
&&\hskip-.28in
\hat{u}_k  = \frac{1}{(2\pi L)^3} \int_{\Om} u(x) e^{-ik\cdot x} \; dx,   \\
&&\hskip-.28in
\hat{v}_k  = \frac{1}{(2\pi L)^3} \int_{\Om} v(x) e^{-ik\cdot x} \; dx,   \\
&&\hskip-.28in
u_k = \sum_{k \leq |j| < 2k} \hat{u}_j e^{ij\cdot x},   \\
&&\hskip-.28in
v_k = \sum_{k \leq |j| < 2k} \hat{v}_j
e^{ij\cdot x},  \\
&&\hskip-.28in
u_k^{<} = \sum_{j < k} u_j,   \qquad v_k^{<} = \sum_{j < k} v_j \\
&&\hskip-.28in u_k^{>} = \sum_{2k \leq j} u_j,  \qquad v_k^{>} =
\sum_{2k \leq j} v_j.
\end{eqnarray*}
The energy balance equation for the Clark$-\alpha$ model for an
eddy of the size $k^{-1}$ is
\begin{eqnarray}
&&\hskip-.28in
\frac{1}{2}\; \frac{d}{dt} (u_k, v_k) +\nu (-\Dd u_k, v_k) = T_k -T_{2k},
\label{BAL}
\end{eqnarray}
where the right hand side is the energy flux and
\begin{eqnarray*}
&&\hskip-.28in T_k  = -\left( B(u_k^{<}, u_k^{<}), v_k  \right) -
\aa^2 \left( B(\nabla u_k^{<}, \nabla u_k^{<}), u_k \right) + \\
&& + \left( B(u_k+u_k^{>}, u_k+u_k^{>}), v_k^{<}  \right) + \aa^2
\left( B(\nabla u_k+\nabla u_k^{>}, \nabla u_k+ \nabla u_k^{>}),
u_k^{<} \right).
\end{eqnarray*}
Taking an ensemble average (long time average) of (\ref{BAL}) we get
\begin{eqnarray}
&&\hskip-.28in
\ang{\nu (-\Dd u_k, v_k)} = \ang{T_k} - \ang{T_{2k}}.
\label{SP-1}
\end{eqnarray}
Let
\[
E_{\aa} (k) = (1+\aa^2 |k|^2) \sum_{|j|=k} |\hat{u}_j|^2.
\]
Then (\ref{SP-1}) can be written as
\[
\nu k^3 E_{\aa} (k)  \sim \int_k^{2k} k^2 E_{\aa} (k) \; dk \sim \ang{T_k} - \ang{T_{2k}}.
\]
For wave numbers $k$ within the inertial range it is assumed that 
there is no energy disspiation, hence we have 
 $\nu k^3 E_{\aa} (k) \ll  \ang{T_k}$, and we also have $\ang{T_k} \approx
\ang{T_{2k}}$, because there is no leakage of energy. One can
consider three possible
scales for the average velocity on an eddy of length size
$k^{-1}$. Namely,
\begin{eqnarray*}
&&\hskip-.28in
 U_k^{(0)}
 = \ang{ \frac{1}{L^3} \int_{\Om} |v_k|^2
dx}^{1/2} \sim \left( k (1+\aa^2 k^2) E_{\aa} (k) \right)^{1/2},   \\
&&\hskip-.28in
 U_k^{(1)} = \ang{ \frac{1}{L^3} \int_{\Om} u_k \cdot v_k
dx}^{1/2} \sim \left( k E_{\aa} (k) \right)^{1/2},   \\
&&\hskip-.28in
 U_k^{(2)} = \ang{ \frac{1}{L^3} \int_{\Om} |u_k|^2
dx}^{1/2} \sim \left( \frac{k E_{\aa} (k) }{ 1+\aa^2 k^2}
\right)^{1/2}.
\end{eqnarray*}
The corresponding turnover time $\tau_k$ for an eddy of the size $k^{-1}$
  will be (cf. \cite{KR67})
\[
\tau_k^{n} \sim \frac{1}{ kU_k^{n}} = \frac{ (1+\aa^2
k^2)^{(n-1)/2}}{k^{3/2} (E_{\aa} (k))^{1/2} }, \qquad n=0, 1, 2.
\]
The energy dissipation rate $\ee$ (\ref{EEE}) is
\[
\ee \sim \frac{1}{\tau_k^{n}} \int_k^{2k} E_{\aa} (k) dk \sim
\frac{k^{5/2} \left( E_{\aa} (k) \right)^{3/2} }{(1+\aa^2
k^2)^{(n-1)/2}}.
\]
As a result, we have
\[
E_{\aa}(k) \sim \frac{\ee^{2/3} (1+\aa^2k^2)^{(n-1)/3}}{k^{5/3}}.
\]
Therefore, the translational kinetic energy spectrum
$E(k)=\sum_{|j|=k}|\hat{u}_j|^2$ is given by
\begin{eqnarray*}
&&\hskip-.28in
E(k)
\equiv
\frac{E_{\aa} (k)}{1+ \aa^2 k^2} \sim \left\{
\begin{array}{ll}   \displaystyle{
\frac{\ee_{\aa}^{2/3}}{k^{5/3}},}
\qquad & \mbox{when  }
k\aa \ll 1\,, \\
\displaystyle{ \frac{\ee_{\aa}^{2/3}}{\aa^{2(4-n)/3}
k^{(13-2n)/3}},} \qquad & \mbox{when  } k\aa \gg 1\,.
\end{array} \right.
\end{eqnarray*}
We refer the reader to \cite{CHOT} and \cite{FHTP} for more details and
discussions of the implication of the energy spectrum
$k^{-(13-2n)/3}$ for the larger wavenumbers ($1 \ll k\alpha$).

\begin{remark}
It is worth stressing that the filtered Clark model does not specify which
one of the $U_k^{(n)}$, for $ n=0,1,2$, is the right average velocity  for
an eddy of the size $k^{-1}$. Consequently, it is not evident what would
be the correct energy spectra in the sub-range, $k\aa \gg 1$, of the
inertial range. This will be a subject of future research. However, our
earlier research suggests the choice $U_k^{(2)}$, for which the
translational energy spectrum for the filtered Clark model has the usual
$k^{-5/3}$ Kolmogorov power law for  wave numbers $k\aa \ll 1$ and
shows $k^{-3}$ decay power law for $k\aa \gg 1.$ 
\end{remark} 

\noindent
\section*{Acknowledgements}
This work was supported in part by the US Department of Energy,
under contract number W--7405--ENG--36 and by Office of Science
ASCAR/AMS/MICS. The work of E.S.T. was supported in part by the
NSF grant No. DMS-0204794, the US CRDF grant No. RM1-2343-MO-02
and by MAOF Fellowship of the Israeli Council of Higher Education.


\begin{thebibliography}{99}
\bibitem{AR75}  R.A. Adams, {\em Sobolev Spaces},
Academic Press, New York, 1975.

\bibitem{CH98} S. Chen, C. Foias, D.D. Holm, E. Olson, E.S. Titi
and S. Wynne,  {\em The Camassa--Holm equations and turbulence.}
 Phys. D {\bf 133} (1999), no. 1-4, 49--65.


\bibitem{CH99} S. Chen, C. Foias, D.D. Holm, E. Olson, E.S. Titi
and S. Wynne,  {\em A connection between the Camassa-Holm
equations and turbulent flows in channels and pipes.}  Phys.
Fluids {\bf 11} (1999), no. 8, 2343--2353.

\bibitem{CH00} S. Chen, C. Foias, D.D. Holm, E. Olson, E.S. Titi
and S. Wynne,  {\em Camassa-Holm equations as a closure model for
turbulent channel and pipe flow.} Phys. Rev. Lett. {\bf  81} (1998), no.
24, 5338--5341.

\bibitem{CH01} S. Chen, D.D. Holm, L.G. Margolin and R. Zhang,
{\em Direct numerical simulations of the Navier--Stokes alpha
model.}  Phys. D {\bf 133} (1999), no. 1-4, 66--83.



\bibitem{CHOT} A. Cheskidov, D.D. Holm, E. Olson and E.S. Titi,  {\em
On a Leray-$\alpha$ model of turbulence,} Proc. Royal Soc. A, (to appear).

\bibitem{CFR79} R.A. Clark, J.H. Ferziger and W.C. Reynolds,
{\em Evaluation of subgrid scale models using an accurately
simulated turbulent flow,} J. Fluid Mech. {\bf 91} (1979), 1--16.

\bibitem{CF85} P. Constantin and C. Foias, {\em Global Lyapunov
exponents, Kaplan--Yorke formulas and the dimension of the
attractors for $2$D Navier--Stokes
equations}, Comm. Pure Appl. Math. {\bf 38} (1985), 1--27.

\bibitem{CF88} P. Constantin and C. Foias, {\em Navier-Stokes Equations,}
  The University of Chicago Press, 1988.

\bibitem{EFNT} A. Eden, C. Foias, B. Nicolaenko and R. Temam, {\em
Exponential Attractors for Dissipative Evolution Equations,}
Research in Applied Mathematics, {\bf 37}, Masson, Paris, 1994.

\bibitem{FHTM} C. Foias, D.D. Holm and E.S. Titi,  {\em
The three dimensional viscous Camassa--Holm equations, and their
relation to the Navier-Stokes equations and turbulence theory,}
J. Dynam. Differential Equations {\bf 14} (2002), 1--35.

\bibitem{FHTP} C. Foias, D.D. Holm and E.S. Titi,  {\em
The Navier--Stokes--alpha model of fluid turbulence.
Advances in nonlinear mathematics and science,} Phys. D {\bf 152/153}
(2001), 505--519.

\bibitem{FOIAS} C. Foias, {\em What do the Navier--Stokes equations
tell us about turbulence?} Harmonic analysis and nonlinear differential  equations
(Riverside, CA, 1995), 151--180, ontemp. Math., {\bf 208}, Amer. Math. Soc.,
Providence, RI, 1997.


\bibitem{FMRT} C. Foias, O. Manley, R. Rosa and R. Temam, {\em
Navier--Stokes Equations and Turbulence,} Cambridge University Press,
Cambridge, 2001.

\bibitem{GA94} G.P. Galdi, {\em An Introduction to the Mathematical
Theory of the Navier-Stokes Equations,} Vol. I \& II,
Springer-Verlag, 1994.

\bibitem{KR67} R.H. Kraichnan, {\em Inertial ranges in two-dimensional turbulence,}
Phys. Fluids {\bf 10} (1967), 1417--1423.

\bibitem{LADY} O.A. Ladyzhenskaya,  {\em The Boundary
Value Problems of Mathematical Physics,} Springer-Verlag, 1985.

\bibitem{LADY91} O.A. Ladyzhenskaya, {\em Attractors for Semigroups
and Evolution Equations,}  Cambridge University Press, Cambridge, 1991.

\bibitem{Le1974}
A. Leonard, {\em Energy cascade in large-eddy simulations of turbulent
fluid flows,} Adv. Geophys. {\bf 18} (1974), 237-.


\bibitem{TT84} R. Temam, {\em Navier-Stokes Equations, Theory and Numerical
Analysis,} 3rd revised edition, North-Holland, 1984.

\bibitem{TT88} R. Temam, {\em Infinite-Dimensional Dynamical Systems in
Mechanics and Physics,} Applied Mathematical Sciences, {\bf 68},
Springer-Verlag, New York, 1988.

\bibitem{VrGeKu1996}
B. Vreman, B. Geurts, and H. Kuerten,
{\em Large-eddy simulation of
the temporal mixing layer using the mixed Clark model,}
Theor. Comput. Fluid Dyn. {\bf 8} (1996), 309-.

\bibitem{VrGeKu1997}
B. Vreman, B. Geurts, and H. Kuerten,
{\em Large-eddy simulation of the
turbulent mixing layer,} J. Fluid Mech. {\bf 339} (1996), 357-.

\bibitem{WWVJ2001} G.S. Winckelmans, A.A. Wray, O.V. Vasilyev, H. Jeanmart,
{\em Explicit-filtering large-eddy simulation using the
tensor-diffusivity model supplemented by a dynamic Smagorinsky term,}
Phys. Fluids {\bf 13} (2001), 1385--1403.
\end{thebibliography}
\end{document}